\documentclass[aip,jcp,amsmath,amssymb,floatfix,reprint,citeautoscript]{revtex4-2}

\usepackage[utf8]{inputenc}
\usepackage{graphicx} 
\usepackage{wrapfig} 
\usepackage[colorlinks,allcolors=black,citecolor=blue,urlcolor=blue]{hyperref} 
\usepackage{ae} 
\usepackage{booktabs} 
\usepackage{silence}
\usepackage{tikz} 
\usepackage{multirow}
\usepackage{chemscheme} 
\usetikzlibrary{matrix} 
\begin{document}

\title{Bridging Electrochemistry and Photoelectron Spectroscopy in the Context of Birch Reduction: Detachment Energies and Redox Potentials of Electron, Dielectron, and Benzene Radical Anion in Liquid Ammonia}

\author{Tatiana Nemirovich}
\thanks{These authors contributed equally}
\affiliation{
Institute of Organic Chemistry and Biochemistry of the Czech Academy of Sciences, Flemingovo nám. 2, 166 10 Prague 6, Czech Republic
}

\author{Vojtech Kostal}
\thanks{These authors contributed equally}
\affiliation{
Institute of Organic Chemistry and Biochemistry of the Czech Academy of Sciences, Flemingovo nám. 2, 166 10 Prague 6, Czech Republic
}

\author{Jakub Copko}
\affiliation{
Institute of Organic Chemistry and Biochemistry of the Czech Academy of Sciences, Flemingovo nám. 2, 166 10 Prague 6, Czech Republic
}

\author{H. Christian Schewe}
\affiliation{
Institute of Organic Chemistry and Biochemistry of the Czech Academy of Sciences, Flemingovo nám. 2, 166 10 Prague 6, Czech Republic
}

\author{Soňa Boháčová}
\affiliation{
Institute of Organic Chemistry and Biochemistry of the Czech Academy of Sciences, Flemingovo nám. 2, 166 10 Prague 6, Czech Republic
}

\author{Tomas Martinek}
\affiliation{
Institute of Organic Chemistry and Biochemistry of the Czech Academy of Sciences, Flemingovo nám. 2, 166 10 Prague 6, Czech Republic
}

\author{Tomas Slanina*}
\email{tomas.slanina@uochb.cas.cz}
\affiliation{
Institute of Organic Chemistry and Biochemistry of the Czech Academy of Sciences, Flemingovo nám. 2, 166 10 Prague 6, Czech Republic
}

\author{Pavel Jungwirth*}
\email{pavel.jungwirth@uochb.cas.cz}
\affiliation{
Institute of Organic Chemistry and Biochemistry of the Czech Academy of Sciences, Flemingovo nám. 2, 166 10 Prague 6, Czech Republic
}

\date{\today}

\begin{abstract}

\setlength\intextsep{0pt}
\begin{wrapfigure}{r}{0.4\textwidth}
  \hspace{-1.8cm}
  \includegraphics[width=\textwidth]{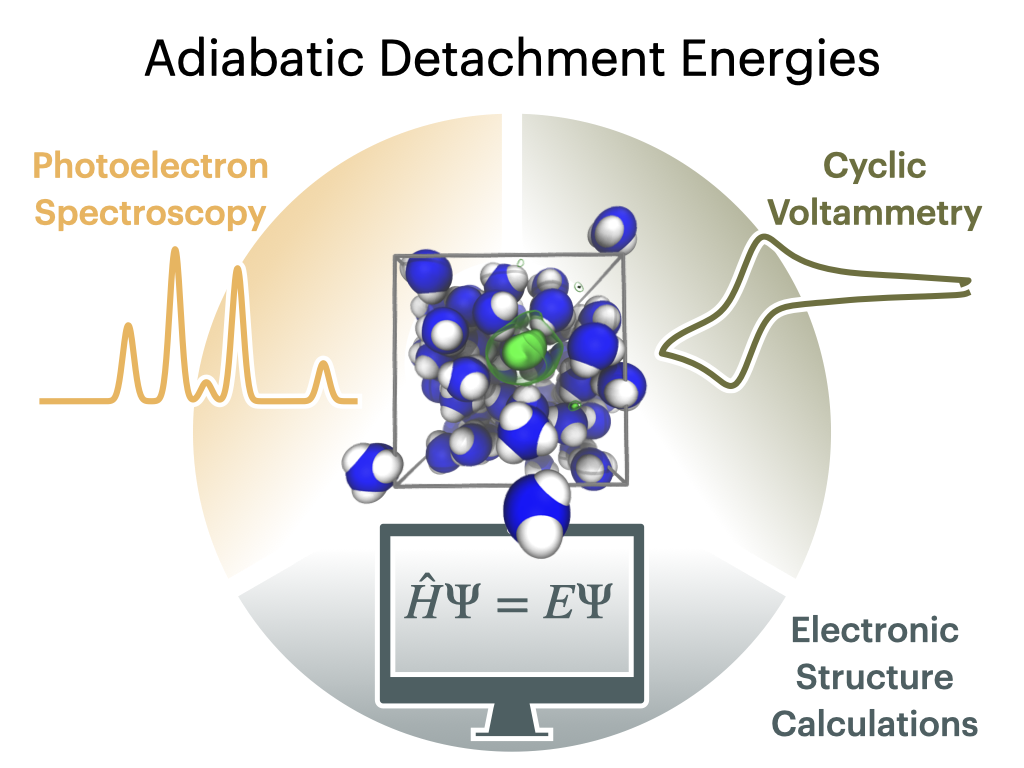}
\end{wrapfigure}

Birch reduction is a time-proven way to hydrogenate aromatic hydrocarbons (such as benzene), which relies on the reducing power of electrons released from alkali metals into liquid ammonia.
We have succeeded to characterize the key intermediates of the Birch reduction process - the solvated electron and dielectron and the benzene radical anion - using cyclic voltammetry and photoelectron spectroscopy, aided by electronic structure calculations.
In this way, we not only quantify the electron binding energies of these species, which are decisive for the mechanism of the reaction but also use Birch reduction as a case study to directly connect the two seemingly unrelated experimental techniques.

\end{abstract}

\maketitle

\section{Introduction}
The experimental techniques of photoelectron spectroscopy~(PES) and cyclic voltammetry~(CV) represent powerful tools for investigating the electronic structure of various chemical systems.
Despite their capabilities to probe essentially the same physical properties, the fields of photoelectron spectroscopy and electrochemistry are distinctly separated with little (if any) mutual overlap, which is also reflected in different languages employed.
In this work, we bridge these two techniques demonstrating the power of such a combination on a case study of chemical species that play a crucial role at an early stage of the Birch reduction process.

Birch reduction is a general tool of organic chemistry employed to regioselectively hydrogenate even the most stable aromatic hydrocarbons by solvated electrons in liquid ammonia in the presence of alcohol as a proton donor~\cite{Birch1946-10.1038/158585c0}.
First, electrons are released into the solvent upon the dissolution of an alkali metal.
This is followed by the formation of solvated electrons that are stable for an extended period of time in liquid ammonia (unlike in water).
Note also that spin-paired dielectrons are formed upon increasing the concentration of the alkali metal~\cite{Zurek2009-10.1002/anie.200900373}.
Second, the generally accepted mechanism of Birch reduction~\cite{Zimmerman2011-10.1021/AR2000698} continues by binding the solvated electron to the benzene molecule forming the benzene radical anion.
All these Birch reduction intermediates -- the solvated electron and dielectron and the benzene radical anion -- are thus negatively charged "electron gain" species.
It is also worth noting that the benzene radical anion is not stable in the gas phase, being a meta-stable shape resonance with a femtosecond life time~\cite{Bazante2015-10.1063/1.4921261, Sanche1973-10.1063/1.1679228}.
In contrast, it is stable enough to be characterized in solution~\cite{Brezina2020-10.1021/acs.jpclett.0c01505, Kostal2021-10.1021/ACS.JPCA.1C04594, Brezina2022-10.1063/5.0076115, Tuttle1958-10.1021/ja01553a005}.
Here we compare the thermodynamic stability of the benzene radical anion with the other species present in the system during the initial steps of Birch reduction, \textit{i.e.}, the solvated electron and dielectron in liquid ammonia, with the aim to find out whether the excess electron prefers to localize on the benzene molecule or rather in a solvent cavity. 
These findings allow us to understand in quantitative terms the capture of solvated electrons by simple arenes.

PES as a high-vacuum technique has been developed originally for measurements of gaseous molecules and low vapor pressure solids.
Later, an extension to aqueous solutions was achieved using the liquid microjet technique~\cite{Winter2006-10.1021/CR040381P/ASSET/IMAGES/LARGE/CR040381PF00032.JPEG} and, most recently, we were also able to probe volatile refrigerated liquids such as liquid ammonia~\cite{Buttersack2019-10.1021/JACS.8B10942, Buttersack2020-10.1126/SCIENCE.AAZ7607}.
In principle, the photoelectron (PE) spectrum carries information about vertical and adiabatic ionization energies at the same time~\cite{Ivanov2018-10.1039/C8CP02936A}.
Note that the absolute value of the adiabatic detachment energy (ADE) is always smaller in absolute terms than the corresponding vertical detachment energy (VDE), with the difference reflecting the amount of nuclear relaxation within the nascent electronic state.
While VDEs are associated directly with the maxima in the spectrum, the determination of ADEs is more complicated.
In the gas phase, ADE is usually assigned with the lowest-lying vibronic spectral feature.
However, in liquids where the vibronic structure tends to be smeared out, one has to rely on the extrapolation of the peak onset towards the baseline~\cite{Wang2000-10.1063/1.481292}.

In comparison, CV is tailored to study thermodynamic and kinetic properties of redox active solutes. 
It reports on electron transfer between analytes and the working electrode by measuring redox potentials extracted from the current response to a time-varying potential applied in cyclic sweeps~\cite{Elgrishi2018-10.1021/ACS.JCHEMED.7B00361/SUPPL_FILE/ED7B00361_SI_002.DOCX}.
Values of the redox potential measurable by CV are limited by electrochemical reduction and oxidation of the employed solvent, \textit{i.e.}, one can only measure potentials within the so-called potential window of the solvent.
This issue of CV renders computational approaches particularly beneficial when interpreting results close to the solvent-specific window edge.

It is straightforward, albeit not fully appreciated, that the ADE from PES and the redox potential from CV refer to the same equilibrium quantity.
Here, we access the ADEs of the solvated electron, the spin-paired dielectron, and the benzene radical anion in liquid ammonia by a novel theoretical approach based on electronic structure calculations and by CV measurements.
Results from both of these approaches are then compared between each other and with previously measured PE spectra.
A one-to-one comparison of ADEs and redox potentials is conditioned by finding a common standard state, which is not guaranteed a priori due to different referencing of the PES and CV measurements.
PES experiments, as well as the corresponding electronic structure calculations, are naturally referred to the vacuum level, while CV data relate to a reference electrode, typically the standard hydrogen electrode~(SHE), and specific experimental conditions.
Attempts to obtain a universal conversion factor between redox potentials with respect to standard electrodes and the vacuum level have been made based on experiments in acetonitrile~\cite{Pavlishchuk2000-10.1016/S0020-1693(99)00407-7} and water~\cite{Ramirez2021-10.1039/D1CP01511G} at room temperature.
However, these values are not directly applicable to the present case of liquid ammonia at low temperatures.
To this end, we employ here a consistent method of conversion between PES and CV data that reflects both the properties of a particular solvent and the temperature dependence.
The present computational and experimental results shed light on thermodynamic stabilities and mutual equilibria of solvated electrons and the benzene radical anion in liquid ammonia with implications for the molecular mechanism of the Birch reduction process.

\section{Methods\label{sec:methods}}

\subsection*{Electronic Structure Calculations}

The double hybrid B2PLYP~\cite{Grimme2006-10.1063/1.2148954} density functional equipped with D3BJ dispersion correction~\cite{Grimme2011-10.1002/JCC.21759, Goerigk2011-10.1021/CT100466K} was employed throughout the study.
Kohn--Sham orbitals were expanded into the Pople triple-$\zeta$ 6-311++g** basis set~\cite{Frisch1998-10.1063/1.447079, Krishnan2008-10.1063/1.438955} augmented by diffuse functions.
Benchmarks of the chosen method and basis set are presented in Section~S1 of the SI. All the electronic structure calculations were carried out using the Gaussian16~\cite{g16A03} software package.


\begin{figure}
    \centering
    \includegraphics[width=\linewidth]{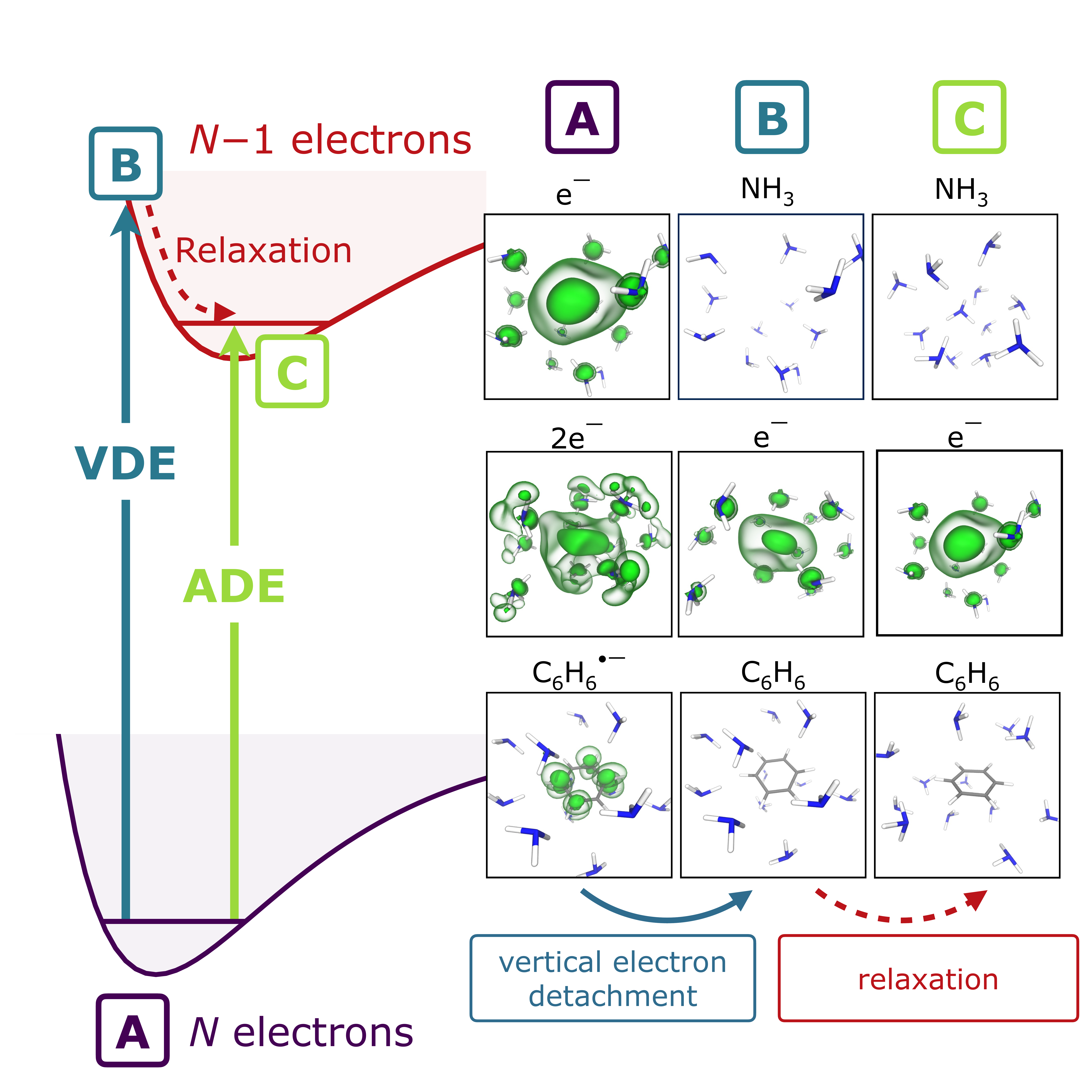}
    \caption
    {
    Left panel: Schematic illustration of vertical (VDE) and adiabatic (ADE) electron detachment process from an initial state having $N$ electron to the final one with $N-1$ electrons.
    Right panel: representatives of structures prior to electron removal (A), directly upon its removal (B), and after relaxation (C) states on the left panel for the three systems. The excess electron(s) is shown as green density contours at two isovalues (opaque at 0.001~\AA$^{-3}$ and transparent at 0.0005~\AA$^{-3}$).  
    }
    \label{fig:VDE-ADE-scheme}
\end{figure}

Both VDEs and ADEs were calculated for the following solutes -- electron, spin-paired dielectron, and benzene radical anion.
The first solvation shell around the solute of twelve solvent molecules was included explicitly by carving molecular clusters out of AIMD trajectories performed by us earlier in Reference~\citenum{Buttersack2019-10.1021/JACS.8B10942}, for neat liquid ammonia, in Reference~\citenum{Buttersack2020-10.1126/SCIENCE.AAZ7607} for the solvated electrons in liquid ammonia and Reference~\citenum{Brezina2020-10.1021/acs.jpclett.0c01505} for the benzene radical anion.
Solvent clusters were centered around the maximally localized Wannier function of the electron and dielectron or the center of mass of the benzene radical anion.
The more distant solvent environment was treated as a structureless medium within the integral equation formalism of the polarizable continuum model (IEF-PCM)~\cite{Cances1998-10.1063/1.474659, Tomasi1999-10.1016/S0166-1280(98)00553-3, Tomasi2005-10.1021/CR9904009} as implemented in Gaussian16~\cite{g16A03}.
Within this approach, liquid ammonia at 223~K was characterized by its experimental values of the total dielectric constant of $\varepsilon=22.6$ or --- for an instantaneous response --- by the high-frequency dielectric constants of $\varepsilon_\infty=1.944$, respectively~\cite{JohnR.Rumble2020}.
A solvent excluded surface (SES) PCM cavity formulation~\cite{Silla1991-10.1002/JCC.540120905, Lange2020-10.1080/00268976.2019.1644384} was employed for all the systems using a smoothing sphere radius of 1.73~\AA.~\footnote{Half of the position of the $1^\mathrm{st}$ peak in the N-N radial distribution function of liquid ammonia; Tomas Martinek (Institute of Organic Chemistry and Biochemistry of the Czech Academy of Sciences, Flemingovo nám. 2, 166 10 Prague 6, Czech Republic), Ondrej Marsalek (Charles University, Faculty of Mathematics and Physics, Ke Karlovu 3, 121 16 Prague 2, Czech Republic), personal communication, August 2021}
This cavity construction approach captures the shape of the cluster and eliminates artifacts arising from the voids between the molecules due to the presence of the solute, as shown in Figure~S2 and previously reported in References~\citenum{Simm2020-10.1002/jcc.26161} and \citenum{Kostal2021-10.1021/ACS.JPCA.1C04594}.

The vertical detachment energy (VDE) was calculated as the difference between the thermally averaged ground-state energies of the system with and without the excess electron at the geometry of the system before the removal of the electron.
Non-equilibrium PCM~\cite{Cossi2000-10.1063/1.480808, Cossi2000-10.1021/JP000997S} (NE-PCM) was employed to capture the vertical character of the process through partial relaxation of the PCM cavity charges due to the high-frequency component of the dielectric constant $\varepsilon_\infty$.

Calculation of the adiabatic detachment energy (ADE) is less straightforward than that of VDE due to the need for accounting also for nuclear relaxation of the system after the electron detachment.
ADE was thus calculated as the difference between the thermal mean ground state energies of the systems with the excess electron and the relaxed system without the excess electron, as obtained from two independent simulations.
The relaxed structures upon electron detachment are the neat liquid ammonia for the solvated electron, the solvated electron for the dielectron, and the solvated benzene molecule for the benzene radical anion.
Note that this approach of calculating ADEs relies on neglecting the entropic contribution to the free energy associated with the nuclear relaxation upon the electron detachment.
In the case of the solvated electron and dielectron, it was shown that entropy correction is below 3~\% of the total contribution to the free energy~\cite{Lepoutre1971-10.1002/BBPC.19710750719}.
This is also in line with our previous study of the structure of the solvated electron and dielectron in liquid ammonia where we report that the organization of the surrounding solvent is very weak~\cite{Buttersack2020-10.1126/SCIENCE.AAZ7607}.
The same assumption was made also for the benzene radical anion~\cite{Brezina2020-10.1021/acs.jpclett.0c01505}.
The calculation protocols for the VDEs and ADEs for all three species are schematically illustrated in Figure~\ref{fig:VDE-ADE-scheme}.

Electronic energies were calculated for 1000 structures of neutral benzene, its radical anion and neat liquid ammonia.
For solvated electron we collected statistics from 200 geometries and for spin-paired dielectron from 170 geometries.
These distributions were constructed as continuous probability densities using the kernel density estimation method as implemented in SciPy~1.9.0~\cite{Virtanen2020-10.1038/s41592-019-0686-2} package using 0.2~eV bandwidth.

\subsection*{Absolute Redox Potential of the SHE}

To determine the solvent-specific conversion factor between the SHE in CV measurements and the vacuum level in PE and theoretical calculations, we have adopted a procedure from Reference~\citenum{Isse2010-10.1021/JP100402X} using the thermodynamic cycle shown in Scheme~\ref{fig:abs-SHE-cycle}.
This approach describes the thermodynamics of the processes occurring at the standard hydrogen electrode and takes care of solvent-specific interactions referencing to the free energy of solvation of a proton, accounting also for its temperature dependence. Similar approaches in the literature~\cite{Trasatti-10.1016/0022-0728(86)80570-8} differ in subtle details resulting in variations of the standard electrode potentials of the order of $\approx$ 0.1~eV. 

Here, we employ the Gibbs free energy of formation of a proton in the gas phase $\Delta G^{(\mathrm{g})}_{\mathrm{H}^+}=1513.32$~kJ/mol~\cite{Isse2010-10.1021/JP100402X} and the Gibbs free energy of solvation of the proton in liquid ammonia at 220~K (temperature used in our experiments) $\Delta G^{(\mathrm{solv})}_{\mathrm{H}^+}=-1219.63$~kJ/mol~\cite{Malloum2017-10.1063/1.4979568} (for more information see the SI), while the Gibbs free energy of the electron in the gas phase is set to zero following the choice of the standard state.
The absolute SHE redox potential then reads
\begin{equation}\label{eq:E_SHE}
    E_\mathrm{SHE}^\mathrm{abs}=\frac{1}{F} \left[\Delta G^{(\mathrm{g})}_{\mathrm{H}^+} + G^{(\mathrm{solv})}_{\mathrm{H}^+} - RT\ln\gamma_\mathrm{H^+} \right] = 3.09\ \mathrm{V}
\end{equation}
where $F$ is the Faraday constant, $R$ is the universal gas constant, $T$ stands for the temperature, and the activity coefficient of a proton in liquid ammonia $\gamma_\mathrm{H^+}=0.2$ is adopted from Reference~\citenum{Lagowski2007-10.1080/15533170601187474}.
Note that the last term in Equation~\ref{eq:E_SHE} corresponds to the change from the Gibbs free energy standard state (the 1~M solution) to the standard state of unit activity.

\begin{scheme}[h]
    \centering
    \begin{tikzpicture}
      \matrix (m) [matrix of math nodes, row sep=1cm, column sep=2cm, minimum width=1.5cm]
      {
         \frac{1}{2} \mathrm{H}_2^\mathrm{(g)} & \mathrm{H}^\mathrm{+ (g)} + \mathrm{e^{- (g)}} \\
         \frac{1}{2} \mathrm{H}_2^\mathrm{(g)} & \mathrm{H}^\mathrm{+ (solv)} + \mathrm{e^{- (g)}} \\};
      \path[-stealth]
        (m-1-1) edge node [left] {$0$} (m-2-1)
                edge node [above] {$\Delta G_{\mathrm{H}^+}^\mathrm{(g)}$} (m-1-2)
        (m-2-1.east|-m-2-2) edge node [below] {}
                node [above] {$F\cdot E_\mathrm{SHE}^\mathrm{abs}$} (m-2-2)
        (m-1-2) edge node [right] {$\Delta G_{\mathrm{H}^+}^\mathrm{(solv)}$} (m-2-2);
    \end{tikzpicture}
    \caption{Thermodynamic cycle used to determine the absolute potential of the SHE.}
    \label{fig:abs-SHE-cycle}
\end{scheme}

\subsection*{Experiments}
\label{sec:methods-experiments}
CV measurements were performed using an Autolab PGSTAT204 potentiostat (Metrohm Autolab B.V.) with computer control (NOVA 2.1.5 software).
A conventional three-electrode configuration consisted of a glassy carbon working electrode (2~mm in diameter), a platinum wire counter electrode, and a silver wire reference electrode.
The working electrode was polished with 1.1~$\mathrm{\mu m}$ alumina and subsequently with a 0.5~$\mathrm{\mu m}$ alumina on a polishing pad.
All CV measurements were performed in a solution of a 0.1~M electrolyte (Bu$_4$NPF$_6$ in dimethoxyethane (DME) at room temperature or KI in liquid ammonia at $-55$~$^{\circ}$C) using the scan rate range from 100 to 1000~mV/s.
Prior to each experiment, the DME solutions were deoxygenated by bubbling nitrogen gas through and the nitrogen atmosphere was maintained throughout the course of the experiment. Liquid ammonia was condensed directly from a cylinder and kept under argon atmosphere during the experiment. 

The CV measurements were referenced internally to our particular setup.
In order to relate our results to the Ag/Ag$^+$ reference electrode, we refered the redox potential of nitrobenzene in liquid ammonia in our setup to the previously measured value~\cite{smith1975electrochemical} (with respect to the Ag/Ag$^+$ electrode).
Therefore, we use a shift of $-0.17$~eV (left part of Scheme~\ref{fig:CV-2-vacuum-shifting}).

Another shift of $+0.59$~V corresponds to the transition from the Ag/Ag$^+$ reference to SHE~\cite{Pavlishchuk2000-10.1016/S0020-1693(99)00407-7}.
Eventually, $+3.09$~V was added to reference the SHE at our experimental conditions to the vacuum level accordingly to Equation~\ref{eq:E_SHE}.
These particular shifts sum up to an overall value of $3.51$~V that must be added to reduction half-wave potential measured in liquid ammonia in our setup at 220~K to reference it to the vacuum (see Scheme~\ref{fig:CV-2-vacuum-shifting}).

\begin{scheme}[b]
    \centering
    \includegraphics[width=\linewidth]{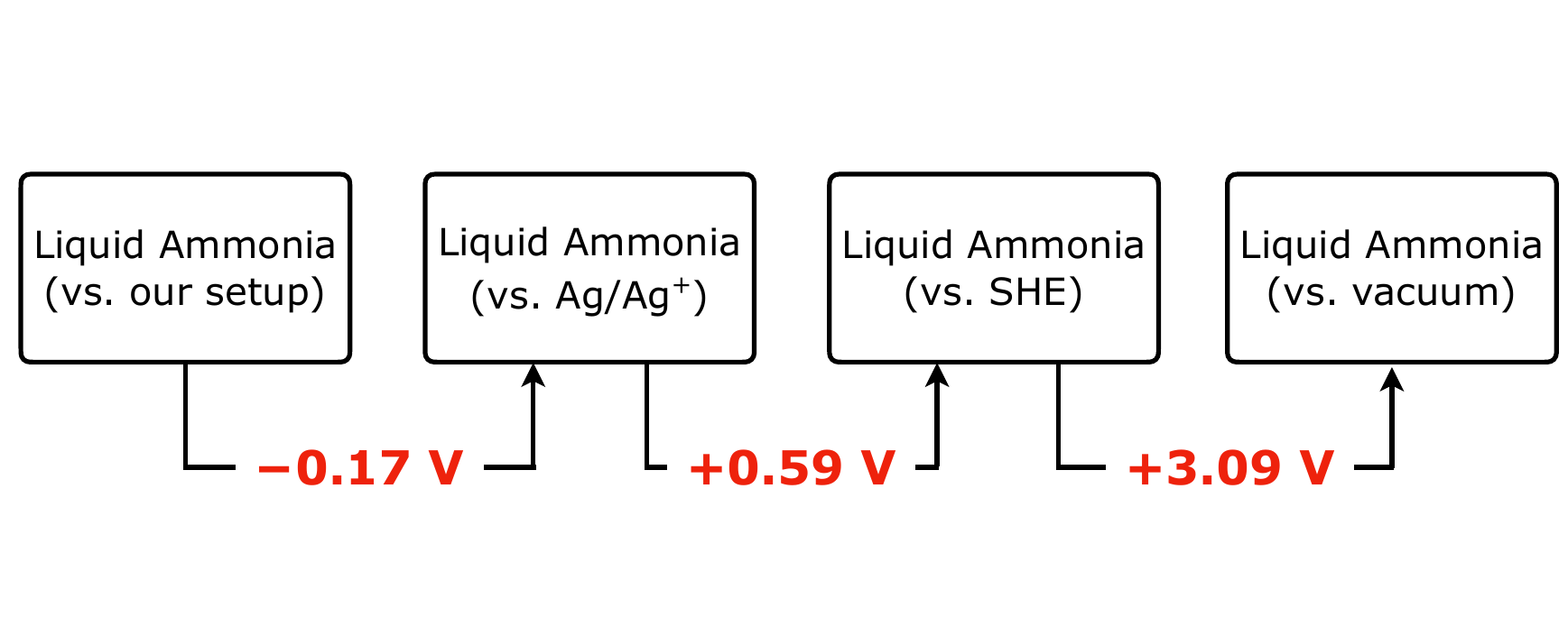}
    \caption{
    Schematic description of shifting the redox potentials measured in liquid ammonia at 220~K referenced to our setup and the vacuum level.
    }
    \label{fig:CV-2-vacuum-shifting}
\end{scheme}

The reduction potential of benzene was found to be out of the range experimentally available in liquid ammonia.
Therefore, we performed the CV measurements in another solvent - dimethoxyethane (DME), where the signal is reachable.
In the next step, to get the redox potential value in liquid ammonia from that in DME, we correlated the reduction peak potentials of several aromatic compounds measurable in both solvents under the same experimental conditions.
Thus for benzene, we first converted the value from DME to that in liquid ammonia and then applied the same shift toward the vacuum level as above.

\section{Results}

\subsection*{Computational Results}

\begin{figure}

    \centering
    \includegraphics[width=\linewidth]{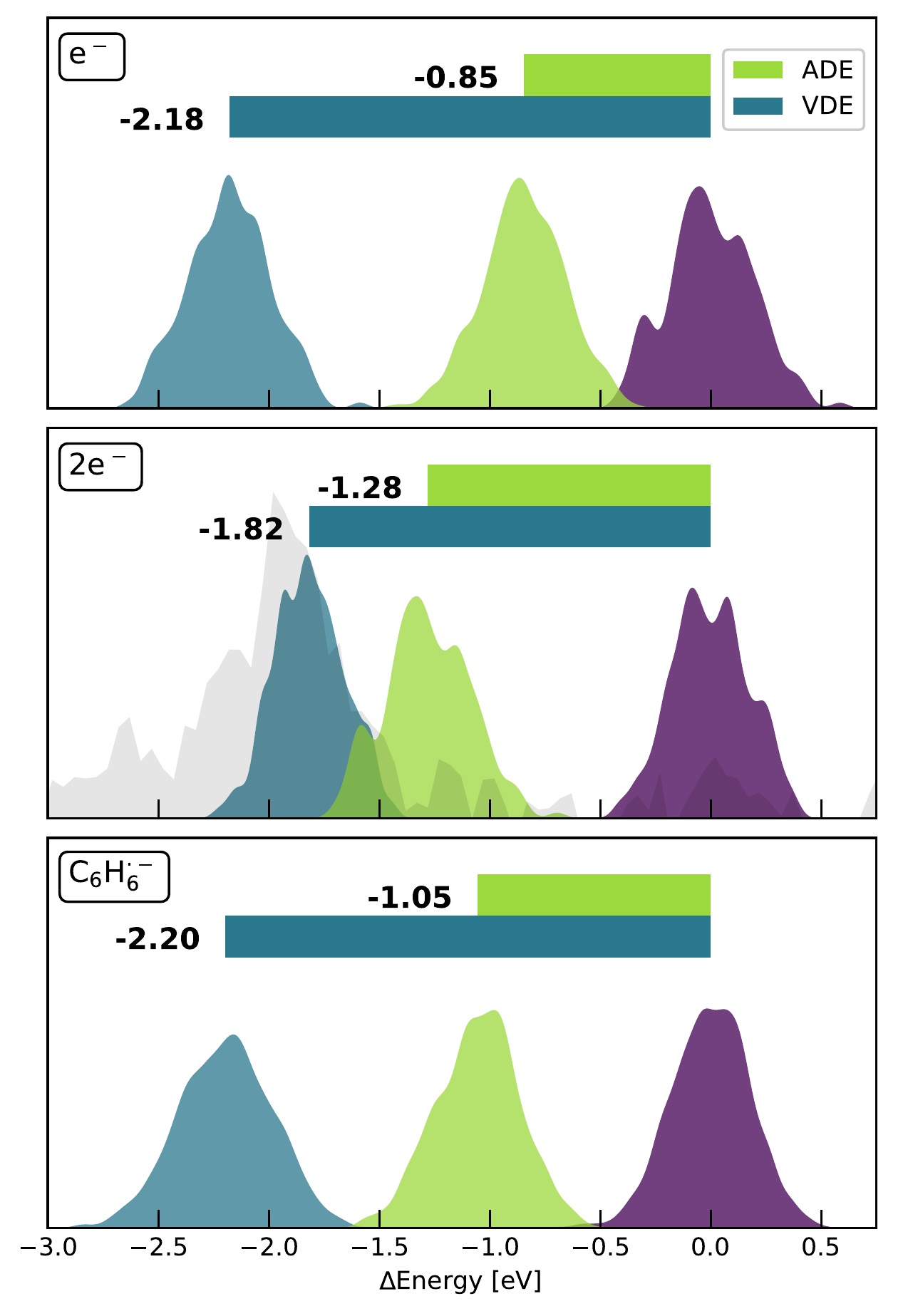}
    \caption
    {
    Normalized energy distributions of the ground-state energies of the systems with the excess electron (purple), ionized structure at unrelaxed geometry (blue), and fully relaxed structures (green).
    Data for the solvated electron are shown on top, for the dielectron in the middle, and for the benzene radical anion in the bottom panel.
    Horizontal bars show the VDEs and ADEs of the respective species.
    The experimental photoelectron spectrum of alkali metal-ammonia solution is shown in gray as published in Figure~3 (0.35~MPM curve) of Reference~\citenum{Buttersack2020-10.1126/SCIENCE.AAZ7607}.
    }
    \label{fig:VDE_ADE_12NH3}
\end{figure}

Distributions of the total electronic ground-state energies of the structures prior to electron removal, directly upon its removal, and after nuclear relaxation are shown in Figure~\ref{fig:VDE_ADE_12NH3} for the solvated electron, spin-paired dielectron, and the benzene radical anion solvated in liquid ammonia.
Solvation was modeled in a hybrid way by including the solute together with the closest 12 solvent molecules explicitly and the bulk solvent was treated as a polarizable continuum (see Section~\ref{sec:methods}).
Benchmark calculations of the VDEs and ADEs with a larger number of explicit solvent molecules, shown in Section~S3 of the SI, demonstrate the convergence of the present hybrid approach to the description of the solvent effects.
All the energies in Figure~\ref{fig:VDE_ADE_12NH3} are aligned with respect to the mean value of the energy distribution (purple) of the species prior to ionization,  which is set to zero.
The resulting values of VDE (blue bar) and ADE (green bar) are then presented on top of the energy distributions in Figure~\ref{fig:VDE_ADE_12NH3}.

The present VDEs of the electron and dielectron are in line with the measured photoelectron spectra in Reference~\citenum{Buttersack2020-10.1126/SCIENCE.AAZ7607}.
Namely, as the experiment was conducted at alkali metal concentrations corresponding to the dielectron regime~\cite{Zurek2009-10.1002/anie.200900373}, we directly compare the experimental spectrum with the VDE of the dielectron (the gray shape in the middle panel of Figure~\ref{fig:VDE_ADE_12NH3}).
The VDE of the benzene radical anion as previously estimated by two independent methods to be $-2.3$~eV ~\cite{Kostal2021-10.1021/ACS.JPCA.1C04594, Brezina2022-10.1063/5.0076115}, is also in line with the present calculated value of $-2.2$~eV.

A value of ADE of $-0.85$~eV was calculated for the solvated electron, while it came as $-1.28$~eV for the dielectron and $-1.05$~eV for the benzene radical anion.
All the ADEs are, in absolute values, smaller than the corresponding VDEs, with the difference being attributed to the relaxation energy of the whole system at the ionized electronic potential energy surface (Figure~\ref{fig:VDE-ADE-scheme}).
From the quantitative point of view, this relaxation is the most pronounced in the case of the solvated electron where the cavity collapses upon ionization yielding a value of the relaxation energy of about 1.3~eV.
When ionizing the dielectron, the resulting solvated electron is of a similar size and still negatively charged thus leaving the cavity radius and orientation of the surrounding solvent molecules rather unchanged.
Consequently, the relaxation energy is just $\approx$ 0.5~eV.
Somewhere in between these two cases, there is the benzene radical anion.
For this species, the cavity in ammonia is formed primarily by the carbon ring while the excess electron makes a relatively small contribution to its size~\cite{Brezina2020-10.1021/acs.jpclett.0c01505}.
Nevertheless, the transition from the benzene radical anion towards neutral benzene is connected with sizable relaxation energy of about 1.1~eV.
Finally, regarding the measured photoelectron spectrum in the middle panel of Figure~\ref{fig:VDE_ADE_12NH3} we note that our calculated value of ADE of the dielectron matches well the onset of the experimental peak.

\subsection*{Experimental Results}

The experimental counterparts to the calculated ADE values are the estimated reduction half-wave potentials obtained by cyclic voltammetry. 
We measured the reduction potential of neat liquid ammonia in our setup described in Section~\ref{sec:methods-experiments} at 220~K as $-2.30$~V that, after shift, resulted in $1.21$~V with respect to the vacuum level (see Section~S4 in SI for the measured voltammogram).
Taking into account that the electron concentration at the surface of the electrode is estimated to be millimolar, solvated dielectrons dominate over monoelectrons~\cite{Zurek2009-10.1002/anie.200900373}.

\begin{figure}[t]
    \centering
    \includegraphics[width=\linewidth]{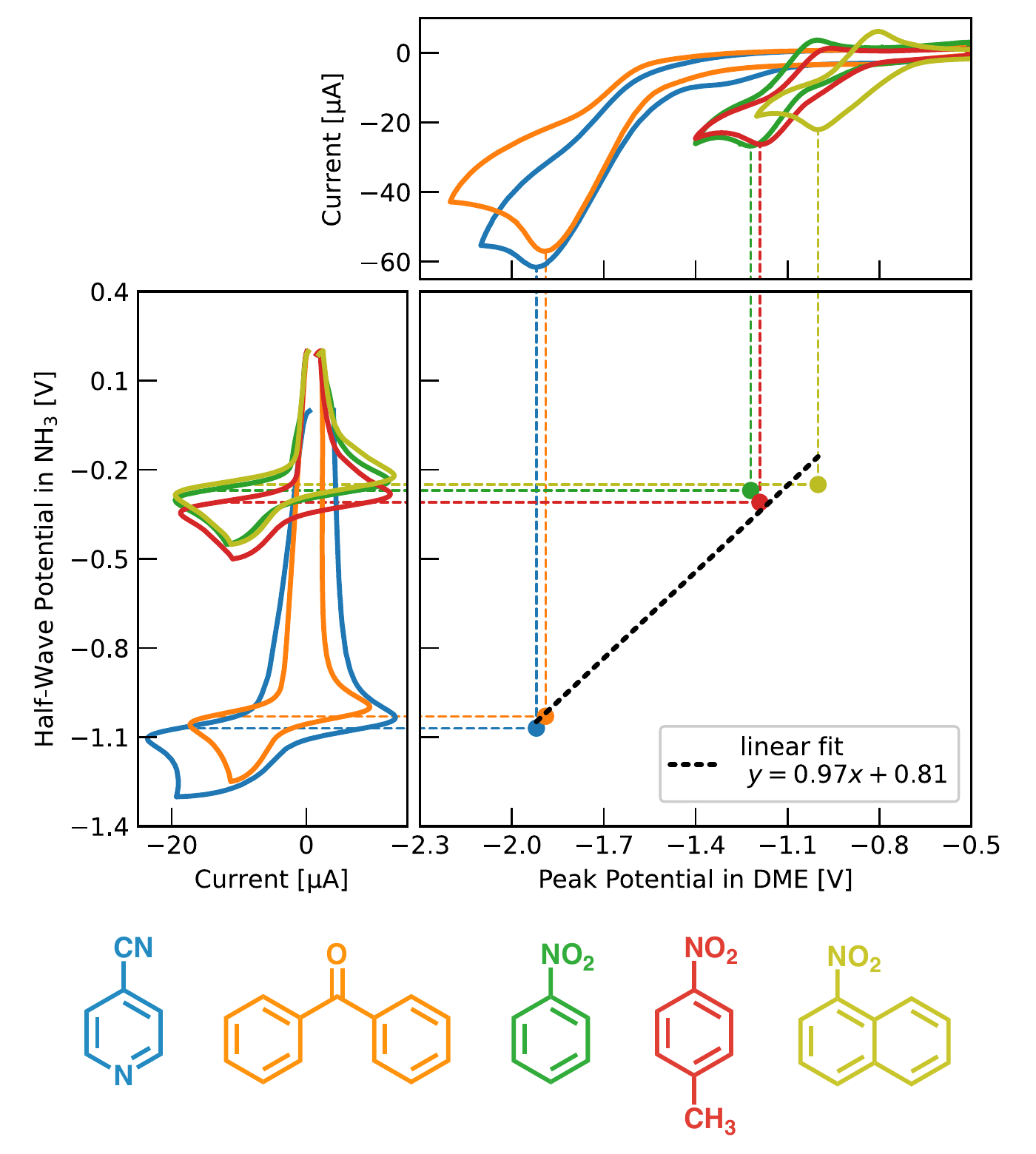}
    \caption{
    Cyclic voltammograms (scan rate 250 mV/s) are plotted for five aromatic compounds in liquid ammonia (left panel) and DME (top panel).
    Correlation of the reduction peak potential in DME with reduction half-wave potential in liquid ammonia vs our setup is shown in the right-bottom panel together with their linear fit including its parameters.
    Also, dashed guidelines connecting the reduction half-wave potential in liquid ammonia and reduction peak potential in DME are provided for each compound.
    Color coding follows as: 4-cyanopyridine --- blue, benzophenone --- orange, nitrobenzene --- green, $p$-nitrotoluene --- red, 1-nitronaphthalene --- olive.
    }
    \label{fig:CV-NH3-DME-correlation}
\end{figure}

In the case of benzene, direct measurement in liquid ammonia appeared to be unfeasible as its reduction potential lies below the reduction point of the solvent.
However, the reduction potential of benzene was reported in DME~\cite{Mortensen1984-10.1002/ANGE.19840960121}.
Therefore, we measured cyclic voltammograms for a series of aromatic compounds both in liquid ammonia and DME as shown in the top and left panel of Figure~\ref{fig:CV-NH3-DME-correlation}.
Note that the selection criteria for the compounds were such that their reduction potentials lie above the reduction potential of the respective solvent and that they are soluble under both conditions.
By comparison of the reduction peak potentials in DME with the reduction half-wave potentials in NH$_3$~(l), we obtained a near-to-linear correlation that is plotted in the right-bottom panel of Figure~\ref{fig:CV-NH3-DME-correlation} (note that similar linear correlation was also found between dimethylformamide and liquid ammonia, see Figure~S4 for details).
Additionally, we purposely avoided any reference system in the measured values in DME as we only aim to estimate the difference between experimental values in DME and liquid ammonia.

\begin{figure}[t]
    \centering
    \includegraphics[width=\linewidth]{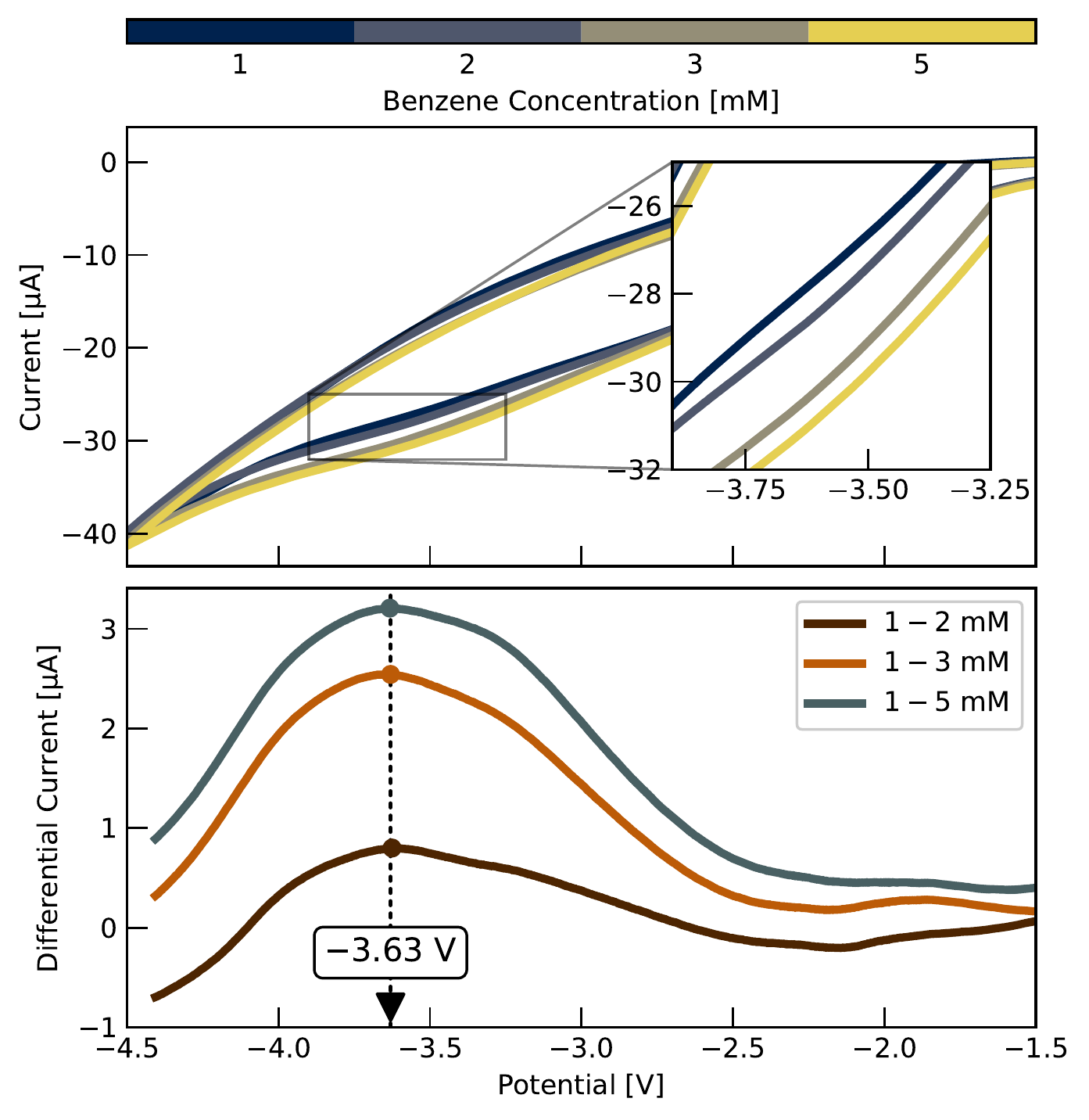}
    \caption{
    CV measurements of benzene in DME at four different concentrations are shown in the top panel together with an inset focusing on the reduction peak of the benzene.
    Three differential curves are shown in the bottom for the four measured concentrations together with their maxima indicated by full circles.
    Mean of the maxima positions at $-3.63$~V is highlighted by the black triangle and dashed line.
    }
    \label{fig:CV-benzene}
\end{figure}

Subsequently, the CV was measured for four DME solutions of benzene of increasing concentration (top panel of Figure~\ref{fig:CV-benzene}).
The reduction of benzene appears close to the window edge of the solvent and thus we performed subtraction of the curves to elucidate the contribution of the benzene reduction to the overall voltammogram.
In particular, the 1~mM curve served as the reference from which the other curves were subtracted.
Then, we identified the reduction peak potential of benzene with the maxima on these differential curves (bottom panel of Figure~\ref{fig:CV-benzene}) that resulted in the mean value of $-3.63$~V.
When referenced to the vacuum level according to Figure~\ref{fig:CV-NH3-DME-correlation} and Scheme~\ref{fig:CV-2-vacuum-shifting}, we obtained the reduction potential of benzene in liquid ammonia of $0.80$~V. In order to directly compare reduction potentials to calculated ADEs we move from V to eV and change sign due to the opposite sign convention when reporting photoelectron energies. The results presented in Table~\ref{tab:exp-theo-results} demonstrate a quantitative agreement between the two approaches.

\begin{table}[h]
    \centering
    \caption{
    Comparison of experimentally measured and theoretically calculated ADEs of electron, dielectron, and benzene radical anion in liquid ammonia.
    The experimental value for electron and dielectron is shared as these two systems cannot be distinguished in the experiment.
    }
    \begin{tabular}{ccc}
         \toprule
         & \multicolumn{2}{c}{ADE [eV]} \\
         \addlinespace[0.25em]
         & Experiment & Calculation\\
         \midrule
         electron & --- & $-0.85$\\
         dielectron & $-1.21$\footnote{Reduction of the neat solvent} & $-1.28$\\
         \multirow{2}{*}{\parbox{3cm}{\centering benzene radical anion}} & \multirow{2}{*}{$-0.80$\footnote{Extrapolated from DME}} & \multirow{2}{*}{$-1.05$}\\
         & & \\
         \bottomrule
    \end{tabular}
    \label{tab:exp-theo-results}
\end{table}

Regarding the measured reduction potential of neat liquid ammonia, it is worth mentioning, that the fact that the reduction potential of benzene lies above the reduction point of the solvent, and theoretical calculations predict that ADE of benzene lies between ADE of solvated electron and dielectron, make strong evidence that the measured value of $-1.21$~V corresponds to the solvated dielectrons.

\section{Discussion}


The development of the liquid microjet technique has allowed, at least in principle, to connect the worlds of photoelectron spectroscopy and cyclic voltammetry.
However, while the peaks in PE spectra report on vertical electron detachment energies without any nuclear relaxation, CV provides via redox potentials information about adiabatic electron detachment energies, which are thermodynamic quantities involving nuclear relaxation.
One way to make a direct connection between the two measurements is to read out the low energy thresholds of the photoelectron peaks.
Provided that the perturbation of the system by photoionization is within the linear response regime and that the resolution of the PES instrument is better than the physical width of the PE peaks, these threshold values indeed report on adiabatic detachment energies~\cite{Buttersack2020-10.1126/SCIENCE.AAZ7607}.
Another way lies in performing electronic structure calculations.
Evaluating the lowest VDE of the system is a rather straightforward procedure --- one simply subtracts energies of the system with and without the excess electron at the geometry pertinent to the situation prior to ionization.
As a matter of fact, for liquid systems one needs to sample a set of thermal geometries of the system with the excess electron, nevertheless, the VDE value converges fast with the number of geometries sampled due to correlated sampling resulting from employing the same geometries before and after photoionization.

This is no more true for the evaluation of ADEs where the geometry after ionization needs to be relaxed, therefore, such calculations are more difficult.
The technique established here, which is based on evaluating the changes in internal energy upon ionization and subsequent nuclear relaxation rely on two assumptions that proved to be valid at least for the present systems.
First, for the evaluation of ADEs, we replaced free energy differences (which are hardly accessible to the electronic structure calculations) with differences in enthalpies (which can be calculated directly for a representative ensemble of system geometries).
This is well justified for the present systems, where the neglected entropic contribution to the free energy amounts to several percents at best~\cite{Lepoutre1971-10.1002/BBPC.19710750719}.
Second, we assumed (and confirmed by benchmark calculations, see SI) that the solutes around the solvated electron/dielectron or the benzene radical anion can be satisfactorily represented by a first layer of explicit solvent molecules embedded in a dielectric continuum of the appropriate dielectric constant.
Using only a limited number of (typically twelve) explicit solvent molecules resulted in rather narrow Gaussian distributions of the internal energies of the system which allowed for an accurate readout of the mean internal energy values.

A key point for a successful connection between PES and CV lies in choosing matching standard states.
This may sound trivial, but in fact, it is not.
PES of non-metallic systems, such as the electrolytes investigated here, is naturally referenced to the vacuum level.
CV redox potential values are, in contrast, reported with respect to a standard electrode.
This is in principle straightforward but in reality, a non-trivial part of the effort lies in connecting the standard states pertinent to the two measurements.
As done here successfully, this means to relate the standard electrode employed in CV to the PES vacuum level accounting for the use of a different solvent at a different temperature than what the standard electrode redox potential value was derived for.   

What does our finding that the reduction potential of the benzene radical anion lies between the reduction potentials of solvated electron and dielectron mean in the context of the Birch reduction process?
It implies that in the low concentration regime (up to units to tens of mM ~\cite{Zurek2009-10.1002/anie.200900373}), where solvated electrons are the prevalent species upon dissolution of alkali metals in liquid ammonia, the formation of the benzene radical anion proceeds as an energetically downhill process.
The situation becomes, however, more complex at higher alkali metal concentrations where solvated dielectrons dominate over solvated electrons. Our results indicate that moving the electron from the dielectron to benzene is connected with a free energy penalty of about 0.2 -- 0.3 eV.
In this context, it is important to realize that while in high concentration regime the initial step of Birch reduction is an uphill process, this is to a large extent compensated by the increased concentration of the reducing agents as compared to the above low concentration regime, making the reaction viable over a broad span of alkali metal concentrations~\cite{greenfield1998kinetics}.
Clearly, several factors -- both thermodynamic and kinetic -- influence the dependence of the efficiency of Birch reduction on the concentration of the solvated electrons and the above free energy consideration is only one of them.

\section{Conclusion}

In this study, we have brought together the fields of photoelectron spectroscopy and cyclic voltammetry using electronic structure calculations as a bridging element.
As a poster child for this approach, we have used the Birch reduction process in liquid ammonia with the key species being the benzene radical anion and the solvated electron and dielectron.
Using CV we have estimated the redox potential of the solvated dielectron and, employing extrapolations from other solvents, also the redox potentials of the benzene radical anion.
Our electronic structure calculations of adiabatic electron detachment energies quantitatively reproduce these redox potential values, providing also the redox potential of the solvated electron (inaccessible to CV).
Bridging toward PES, the present calculations also reproduce well the spectroscopically determined vertical detachment energies.

\section*{Supporting Information}

Detailed methodological benchmarks, comparison data for larger explicit clusters, visualization of the PCM cavities.
Cyclic voltammograms measured at different scan rates as well as additional correlation of redox potentials in different solvents.
Structures employed in the electronic structure calculations together with appropriate input files are included.

\section*{Acknowledgement}

P.J. is thankful for support from the European Regional Development Fund (Project ChemBioDrug no. CZ.02.1.01/0.0/0.0/16${\_}$019/0000729). 
T.N. and T.M. acknowledge support from University of Chemistry and Technology Prague where they are enrolled as PhD. students.
V.K. acknowledges support from Faculty of Science, Charles University where he is enrolled as a PhD. student.
T.N. and V.K. also acknowledge support from the IMPRS for Many Particle Systems in Structured Environments.
T.S. acknowledges the INTER-COST grant (No. LTC20076) provided by the Czech Ministry of Education, Youth and Sports.
Computational resources were supplied by the project "e-Infrastruktura CZ" (e-INFRA CZ LM2018140) supported by the Ministry of Education, Youth and Sports of the Czech Republic.

\section*{References}

%

\end{document}


\title{Supporting Information for: Bridging Electrochemistry and Photoelectron Spectroscopy in the Context of Birch Reduction: Detachment Energies and Redox Potentials of Electron, Dielectron, and Benzene Radical Anion in Liquid Ammonia}

\author{Tatiana Nemirovich}
\thanks{These authors contributed equally}
\affiliation{
Institute of Organic Chemistry and Biochemistry of the Czech Academy of Sciences, Flemingovo nám. 2, 166 10 Prague 6, Czech Republic
}

\author{Vojtech Kostal}
\thanks{These authors contributed equally}
\affiliation{
Institute of Organic Chemistry and Biochemistry of the Czech Academy of Sciences, Flemingovo nám. 2, 166 10 Prague 6, Czech Republic
}

\author{Jakub Copko}
\affiliation{
Institute of Organic Chemistry and Biochemistry of the Czech Academy of Sciences, Flemingovo nám. 2, 166 10 Prague 6, Czech Republic
}

\author{H. Christian Schewe}
\affiliation{
Institute of Organic Chemistry and Biochemistry of the Czech Academy of Sciences, Flemingovo nám. 2, 166 10 Prague 6, Czech Republic
}

\author{Soňa Boháčová}
\affiliation{
Institute of Organic Chemistry and Biochemistry of the Czech Academy of Sciences, Flemingovo nám. 2, 166 10 Prague 6, Czech Republic
}

\author{Tomas Martinek}
\affiliation{
Institute of Organic Chemistry and Biochemistry of the Czech Academy of Sciences, Flemingovo nám. 2, 166 10 Prague 6, Czech Republic
}

\author{Tomas Slanina*}
\email{tomas.slanina@uochb.cas.cz}
\affiliation{
Institute of Organic Chemistry and Biochemistry of the Czech Academy of Sciences, Flemingovo nám. 2, 166 10 Prague 6, Czech Republic
}

\author{Pavel Jungwirth*}
\email{pavel.jungwirth@uochb.cas.cz}
\affiliation{
Institute of Organic Chemistry and Biochemistry of the Czech Academy of Sciences, Flemingovo nám. 2, 166 10 Prague 6, Czech Republic
}

\date{\today}

\begin{abstract}
    
\end{abstract}

\maketitle

\section{Electronic structure benchmarks}

First, the convergence of the results with respect to the method and basis set of choice is rationalized in the following paragraphs.
Second, we show additional computational results using a larger explicit region around the solvated species.

\subsection*{Method \& Basis Set}

\begin{figure*}
    \centering
    \includegraphics[width=0.90\linewidth]{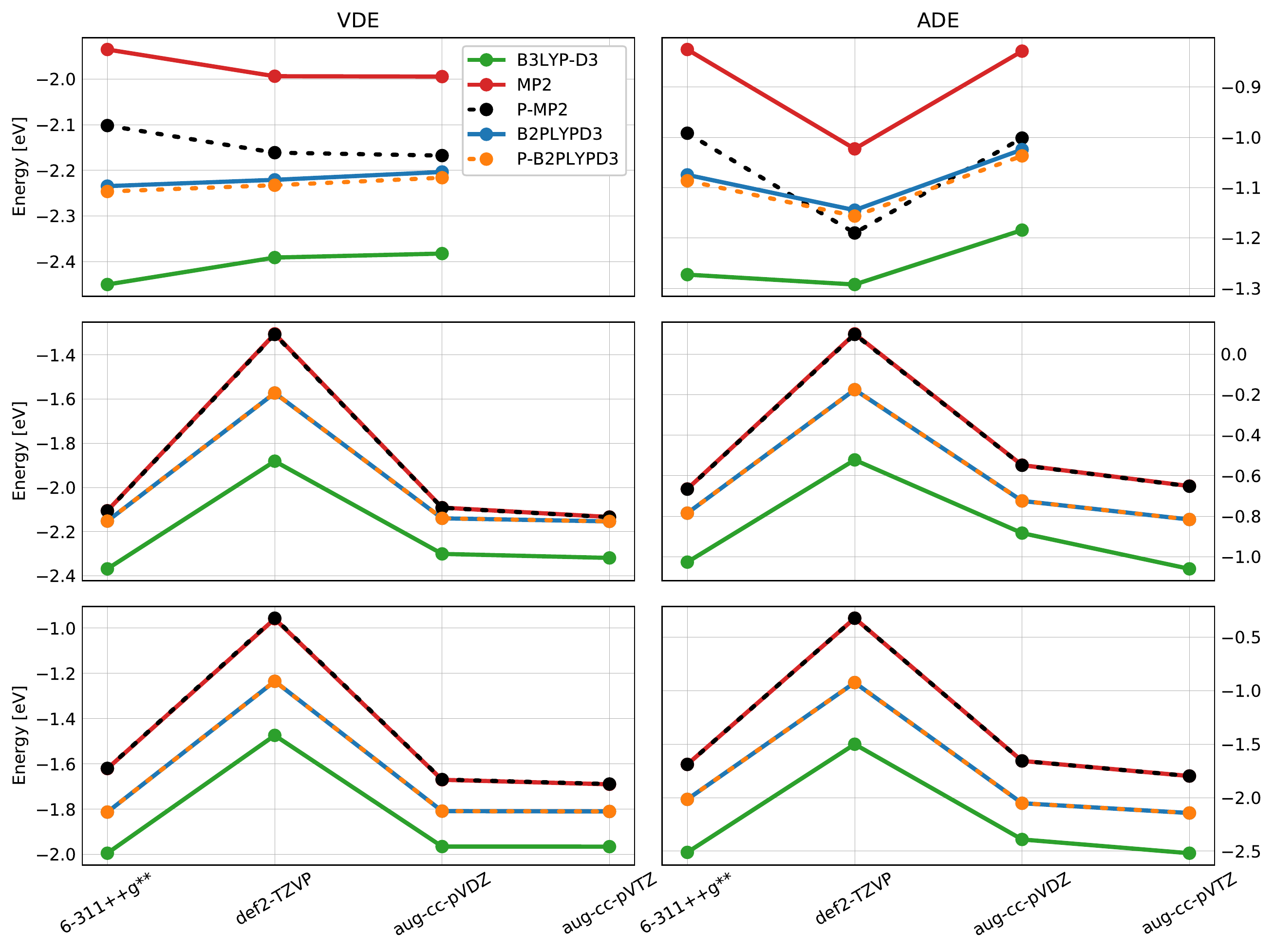}
    \caption
    {
    Benchmarks of method and basis set choice for calculating VDEs and ADEs for the benzene radical anion in the top row, the electron in the middle row, and the spin-paired dielectron in the bottom row.
    Mean values of 10-20 independent structures are shown.
    }
    \label{fig:method-basis-benchmark}
\end{figure*}

We tested the performance of several electronic structure methods together with different basis sets for calculating both VDEs and ADEs on a small batch of uncorrelated structures.
Namely, we employed the M{\o}ller Plesset second-order perturbation theory (MP2), hybrid B3LYP, and double hybrid B2PLYP density functionals.
Latter two were equipped with the D3BJ dispersion correction.
These methods were combined with some common basis sets: Pople triple-$\zeta$ 6-311++g** augmented by diffuse functions, Karlsruhe Def2-TZVP, and correlation consistent aug-cc-pVDZ and aug-cc-pVTZ.
These particular choices were made according to our previous experience with performing AIMD simulations and VDEs calculations of the species in question in this work.
Results of the benchmarks are shown in Figure~\ref{fig:method-basis-benchmark} for the benzene radical anion, electron, and spin-paired dielectron.

The average of both VDEs and ADEs of the benzene radical anion (top row of Figure~\ref{fig:method-basis-benchmark}) appear to be only slightly affected by the basis set choice.
However, we observe a significant discrepancy between the B3LYP and the MP2.
We attribute it mainly to the large higher spin state contamination of the Kohn--Sham wavefunction when performing the spin-polarized calculation.
The expectation value of spin-squared operator $\hat{S}^2$ is usually of $\approx0.85$ when using the MP2 method which deviates significantly from the correct value of $S^2=S(S+1)=0.75$ for $S=0.5$.
This issue was improved by the annihilation of the higher-spin components within the projection method (P-MP2) as described in Reference~\citenum{Schlegel199810.1063/1.450026}; compare black dashed line and full red lines in the top panels of Figure~\ref{fig:method-basis-benchmark}.
In the case of the solvated electron and dielectron, similar discrepancies between B3LYP and MP2 method were observed, however without spin contamination and thus no effect when P-MP2 was applied.
The best of the two approaches we find in the double hybrid DFT; B2PLYPD3 in particular.
It employs the Kohn--Sham orbitals as the regular hybrid DFT, thus the method does not suffer from the spin-contamination for the benzene radical anion while accounting for the perturbative MP2-like correction at the same time.

Also, the basis set choice was tested.
The results for the benzene radical anion seem to be only slightly dependent on augmentation of the basis set due to the close arrangement of the benzene atoms which is in line with our previous study~\cite{Kostal2021-10.1021/ACS.JPCA.1C04594} where the def2-TZVP basis set was used.
On the other hand, missing nuclei in the case of solvated electron and dielectron make use of diffuse functions necessary for their correct description as we show in the two bottom panels in Figure~\ref{fig:method-basis-benchmark}.
We show that the smaller, thus cheaper triple-$\zeta$ 6-311++g** basis set produces very similar results as the larger correlation consistent bases.

Therefore we used the B2PLYPD3/6-311++g** combination of electronic structure method and basis set as the method of choice consistently throughout this study for all the systems.

\newpage
\section{PCM cavity details}
\begin{figure}[h]
    \centering
    \includegraphics[width=\linewidth]{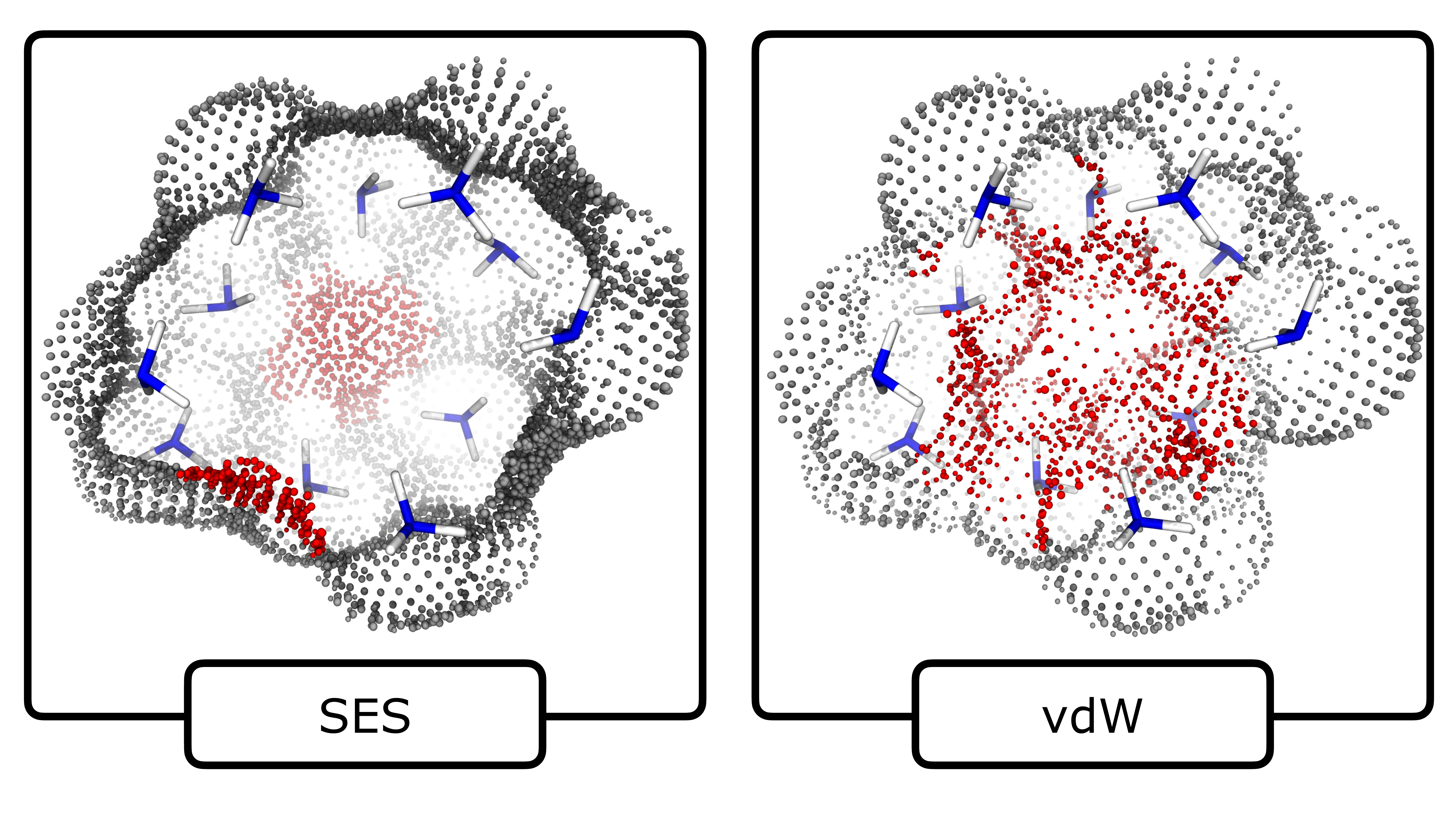}
    \caption
    {
    An ammonia cluster surrounded by the PCM point charges containing a cavity due to the solvated electron in its center.
    The vdW surface is shown on the right and the SES on the left.
    Spurious cavity points are colored in red.
    }
    \label{fig:PCM-cavities}
\end{figure}

In this section, we show a visual comparison of the two generally accepted formulations of the PCM cavity: the van der Waals (vdW) which is the default in Gaussian16, and solvent excluded surface (SES).
Both approaches use the overlapping vdW spheres, however, the latter method employs a smoothing procedure; a probe sphere that "rolls" on the original vdW surface and parts where the probe cannot fit are replaced by a smooth surface.
It results into the complete removal of those points which are artificially between molecules.
Note that these two approaches are identical for small molecules but differ significantly for molecular clusters.
In Figure~\ref{fig:PCM-cavities} we show these two surfaces for one ammonia cluster with a void in its center due to the solvated electron.
The cavity points spuriously spilled between the ammonia molecules in the vdW cavity are shown in red.
For this distinction, we use a threshold of 3.9~{\AA} that corresponds to the position of the peak of the solvated electron - ammonia nitrogen radial distribution function.
The cavity surface is found unphysically spilled into the center of the cluster in the case of the simpler vdW formulation due to missing nuclei in its center.
Note that the red points are also found for the SES but here they do not spill between the molecules but rather create a surface due to missing explicit solvent at the edge of the cluster.

\section{Larger Explicit Clusters}

\begin{figure}[b]
    \centering
    \includegraphics[width=0.95\linewidth]{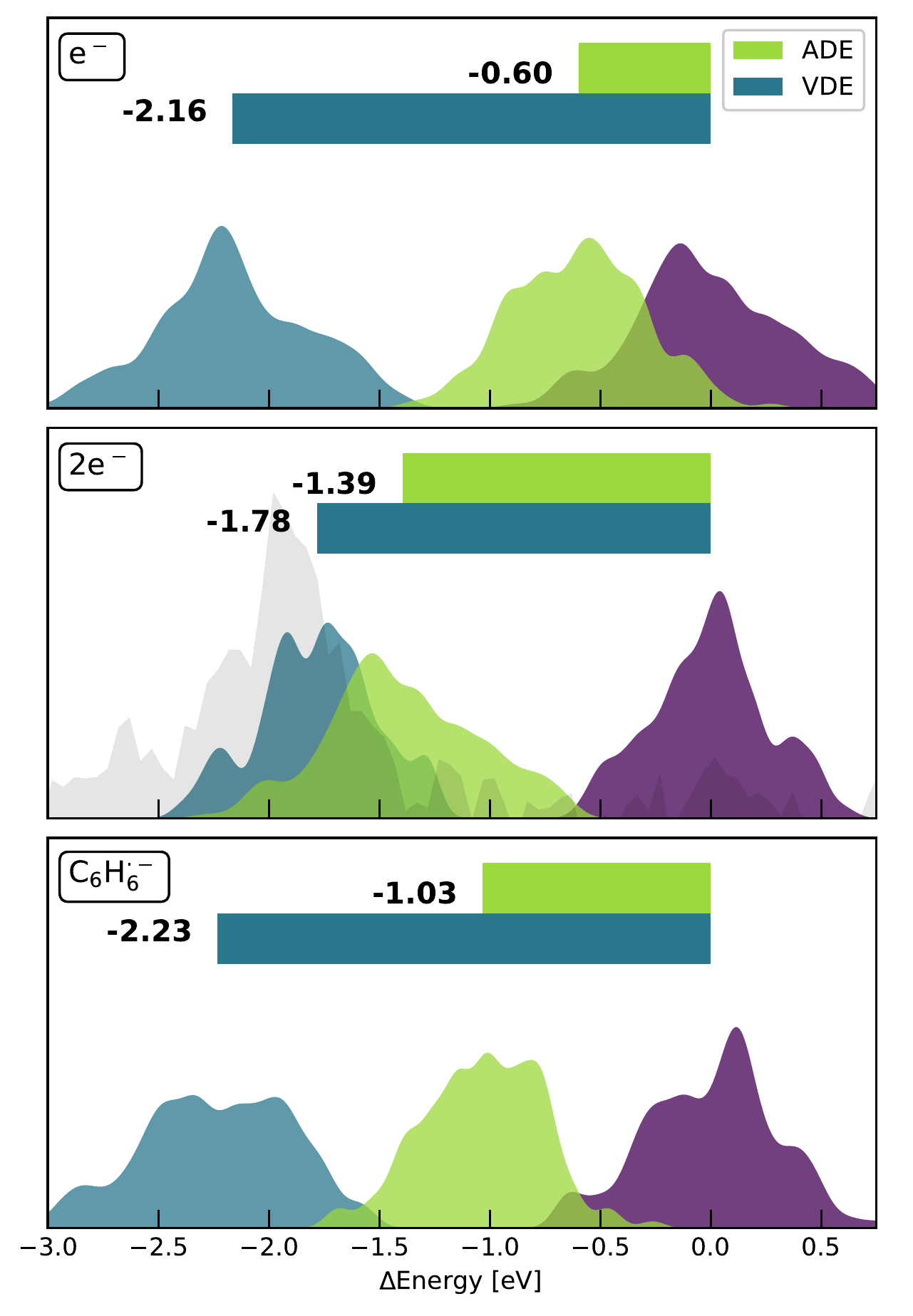}
    \caption
    {
    Normalized energy distributions of the ground-state energies of the systems with the excess electron (purple), the ionized structure at unchanged geometry (blue), and fully relaxed structures (green).
    Each species is solvated by 24 explicit ammonia molecules.
    Data for the solvated electron are shown on top, dielectron in the middle, and the benzene radical anion in the bottom panel.
    Horizontal bars show the VDEs and ADEs of the respective species.
    The experimental photoelectron spectrum is shown in gray as published in Figure~3 (0.35~MPM curve) of Reference~\citenum{Buttersack2020-10.1126/SCIENCE.AAZ7607}.
    }
    \label{fig:VDE-ADE-24NH3}
\end{figure}

Here we present VDEs and ADEs of all three species studied in this work using 24 explicit ammonia molecules for comparison to the results in the main text, where 12 ammonia were employed.
A sample of 200 structures was used to calculate VDEs and ADEs.
From the results shown in Figure~\ref{fig:VDE-ADE-24NH3} we can see a broadening of all energy distributions due to the larger space of accessible geometries of the clusters.
Here, the distributions were constructed by the same method as described in the main text in Section~3 (Computational Results).
Moreover, we see quantitative changes in the VDEs and ADEs of the solvated electron and dielectron of about 0.2~eV while results for the solvated benzene radical anion remained practically unaffected.
Note that the relative positions of the VDEs and ADEs unchanged with the larger explicit solvent region.

In this work we follow methods of applying Marcus theory of electron transfer (ET) within the
framework of PES~\cite{Wang2000-10.1063/1.481292, Ghosh2012-10.1021/JP301925K, Schroeder2015-10.1021/JA508149E}. There, the photodetachment process is
interpreted as an oxidation process analogous to half of an electron transfer (ET) reaction – since
there is no electron acceptor the second half of ET is obsolete. Thus, the difference between the
vertical and adiabatic detachment energies is equivalent to reorganizational energy.

\newpage

\section{Cyclic Voltammetry Measurements}

\begin{figure}[h]
    \centering
    \includegraphics[width=\linewidth]{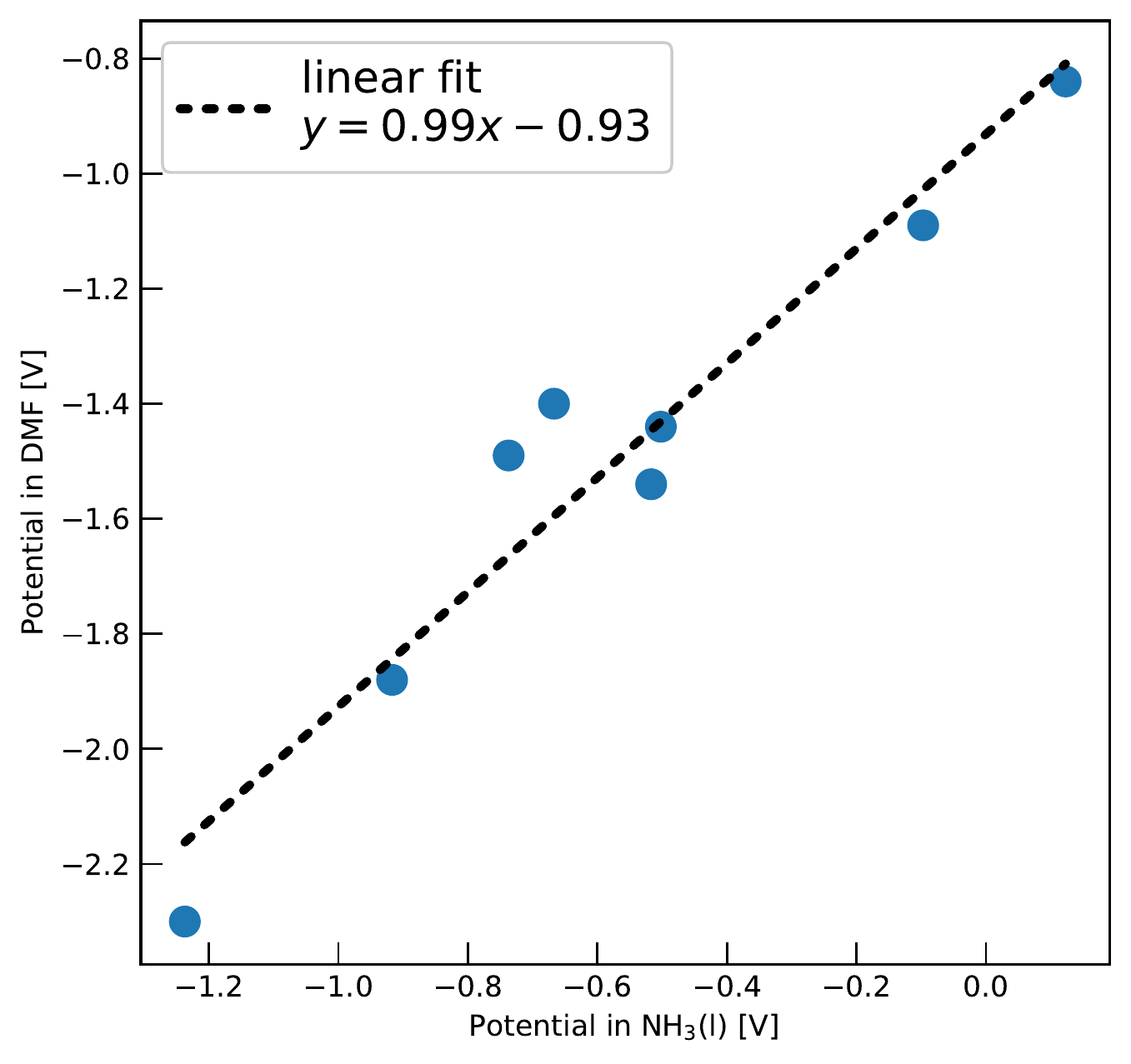}
    \caption{Correlation of the measured half-wave potentials in liquid ammonia ($x$-axis) and in DMF ($y$-axis).
    A linear fit is provided by the dashed black line together with its parameters.
    Individual values together with their references are shown in Table~\ref{tab:exp-theo-results}.}
    \label{fig:NH3-DMF-correlation}
\end{figure}

In this section we take from the literature the redox potentials of different organic compounds in other solvents - namely dimethylformamide (DMF) and liquid ammonia, and show the correlation between the redox potentials is also linear (Figure~\ref{fig:NH3-DMF-correlation}).

\begin{table}[h]
    \caption{Published values of redox potentials vs. SCE of studied compounds in liquid ammonia and in DMF.}
    \centering
    \begin{tabular}{rcc}
    \toprule
     & \multicolumn{2}{c}{Reduction Potential [V]} \\
    \cmidrule{2-3} 
     & NH$_3$(l) & DMF \\
     \midrule
     azobenzene & $-0.502$\cite{Herlem2006-10.1080/00032718008077686} & $-1.440$\cite{goulet2017electrocatalytic} \\
     pyridazine & $-1.237$\cite{Herlem2006-10.1080/00032718008077686} & $-2.300$\cite{Maruyama1979-10.1016/S0022-0728(79)80393-9} \\
     cinnoline & $-0.737$\cite{Herlem2006-10.1080/00032718008077686} & $-1.490$\cite{degrand1980} \\
     benzocinnoline & $-0.667$\cite{Herlem2006-10.1080/00032718008077686} & $-1.400$\cite{degrand1980} \\
     nitrobenzene & $-0.097$\cite{smith1975electrochemical} & $-1.090$\cite{Bock1985-10.1515/znb-1985-1108} \\
     nitrosobenzene & $0.123$\cite{smith1975electrochemical} & $-0.840$\cite{williot1999addition} \\
     cinnamonitrile & $-0.917$\cite{vartires1975} & $-1.880$\cite{Ohno1996-10.1016/0040-4020(95)01018-1} \\
     diethylfumarate & $-0.517$\cite{vartires1975} & $-1.540$\cite{baizer1972} \\
     \bottomrule
    \end{tabular}
    \label{tab:exp-theo-results}
\end{table}

\begin{figure*}[t]
    \centering
    \includegraphics[width=0.9\linewidth]{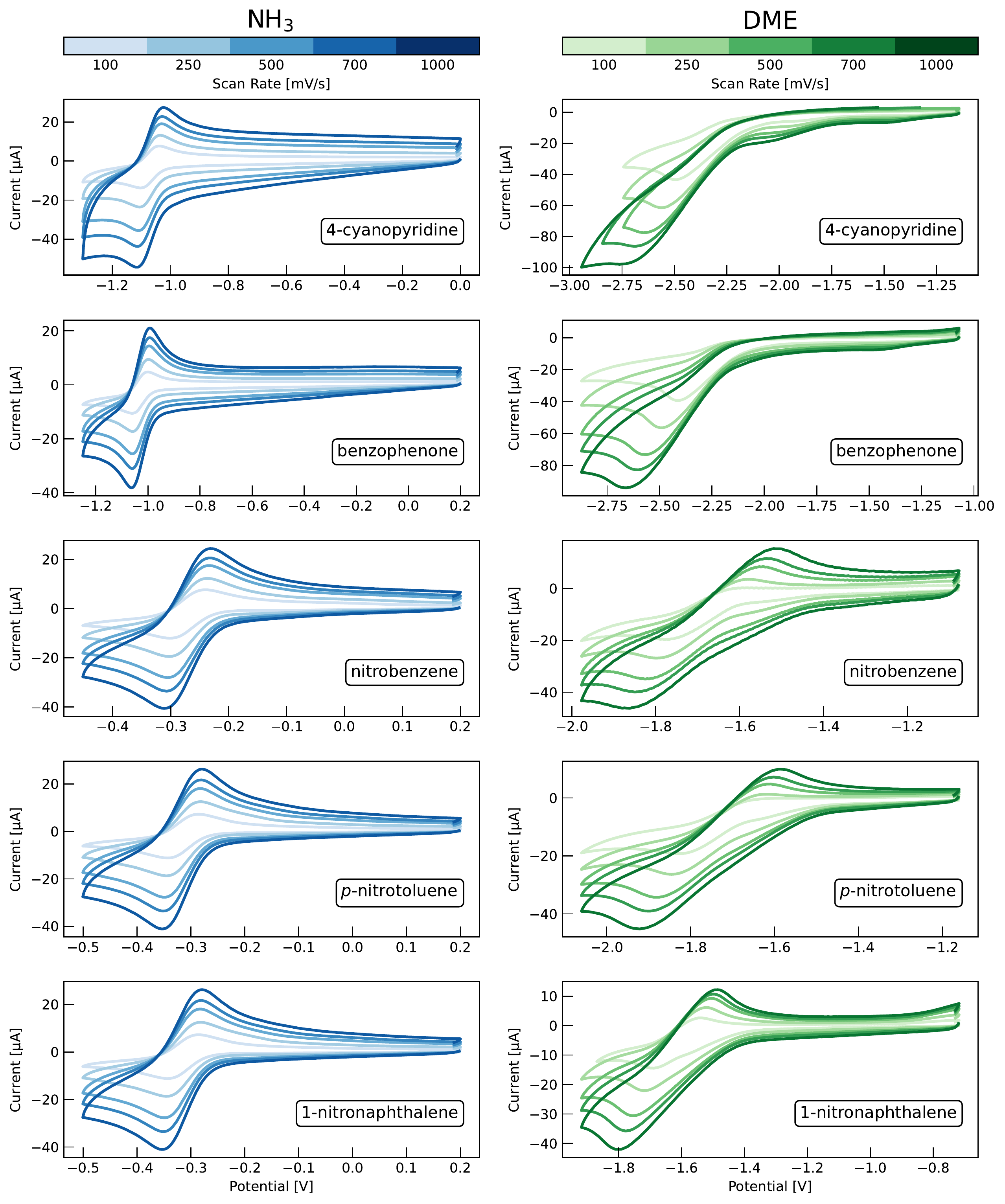}
    \caption{Cyclic voltammograms of measured compounds (1 mM) with KI (0.1 M) in liquid ammonia (left panel) and Bu$_4$NPF$_6$ (0.1 M) in DME (right panel) at different scan rates (100-1000 mV/s).}
    \label{fig:voltammograms}
\end{figure*}

Also, cyclic voltammograms for the five correlation compounds measured at different scan rates in DME and liquid ammonia are presented (Figure~\ref{fig:voltammograms}).

\clearpage

\begin{figure}
    \centering
    \includegraphics[width=\linewidth]{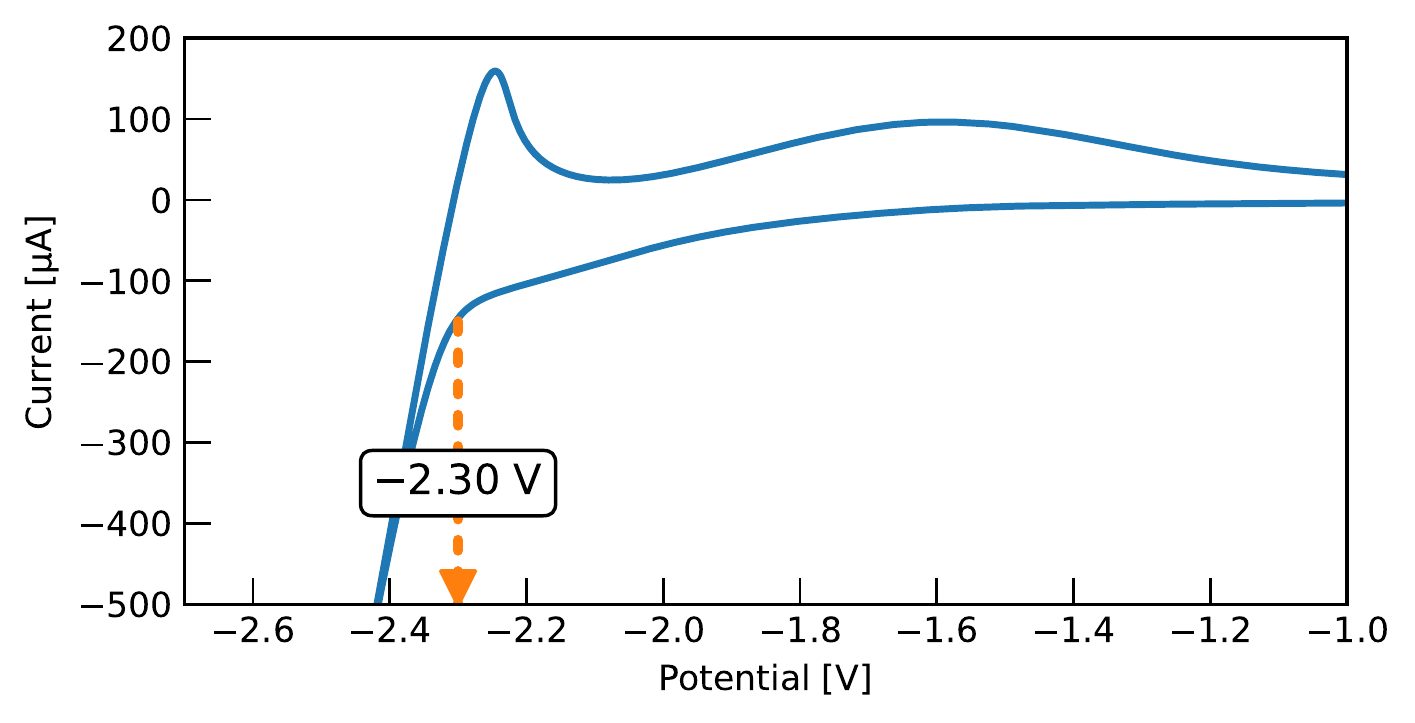}
    \caption{Cyclic voltammogram of neat liquid ammonia at -55°C with KI (0.1~M) at scan rate 250~mV/s.}
    \label{fig:CV-NH-neat}
\end{figure}

In Figure~\ref{fig:CV-NH-neat}, the cyclic voltammogram of the neat liquid ammonia is shown and the reduction point of the solvent itself is indicated by an orange dashed line with an arrow. 
This corresponds to generation of solvated dielectrons.

\begin{figure}
    \centering
    \includegraphics[width=\linewidth]{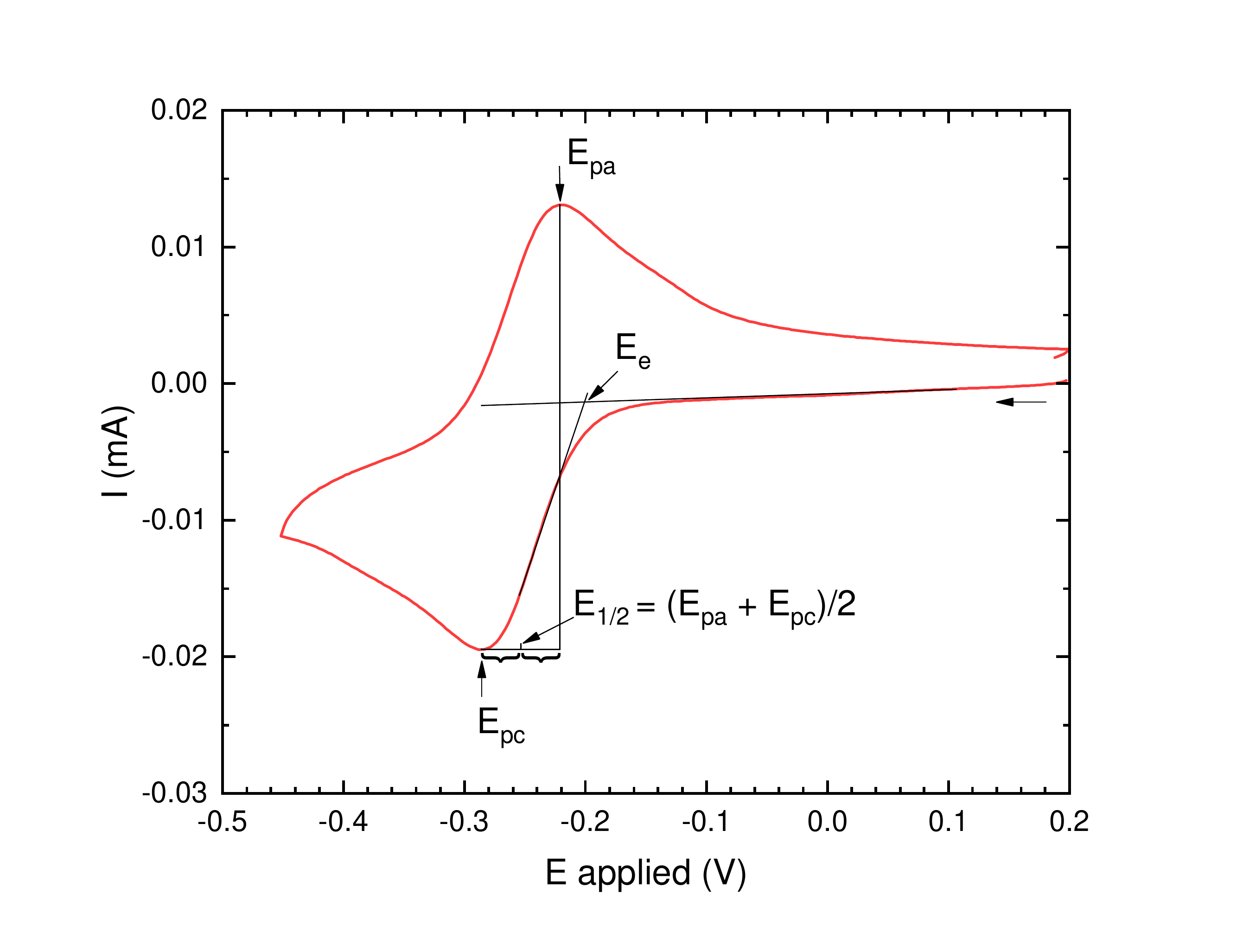}
    \caption{Evaluation of peak potentials $E_\mathrm{p}$ (c-cathodic, a-anodic), half-wave potentials $E_{1/2}$ and onset potentials $E_\mathrm{e}$ from cyclic voltammograms.}
    \label{fig:CV-peak-pot-eval}
\end{figure}

\begin{figure}
    \centering
    \includegraphics[width=\linewidth]{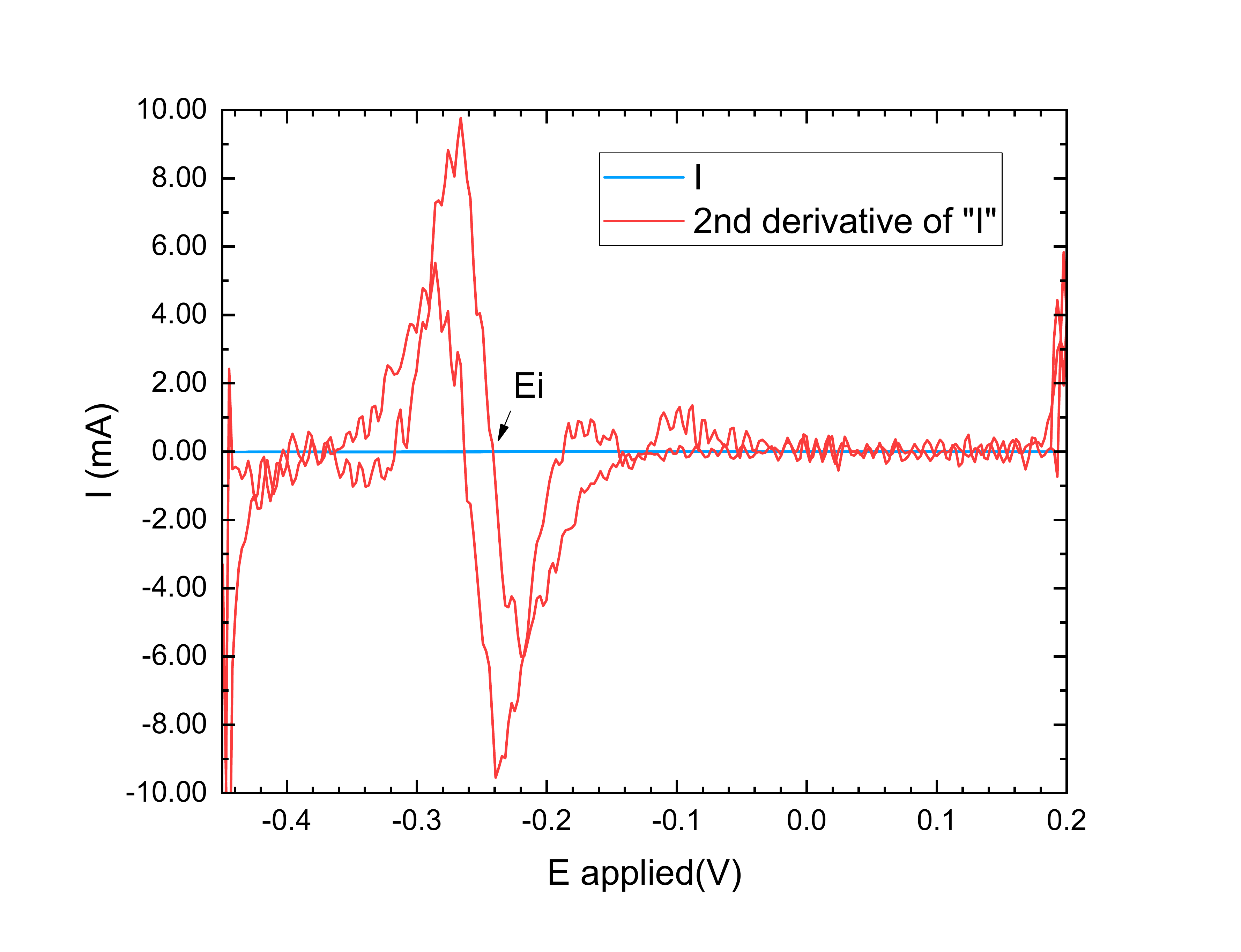}
    \caption{Evaluation of inflection potentials $E_\mathrm{i}$ from cyclic voltammograms.}
    \label{fig:CV-inflection}
\end{figure}

Evaluation of voltamogramms is shown in Figures~\ref{fig:CV-peak-pot-eval} and~\ref{fig:CV-inflection} and obtained values are shown in Table~\ref{tab:CV-potentials}.
However, only reduction peak potential ($E_\mathrm{pc}$) in DME and reduction half-wave potential ($E_\mathrm{1/2}$) in liquid ammonia are used for correlation purposes.

\begin{table}[ht]
    \centering
    \caption{Values of the cathodic, inflection and onset potentials for the five different aromatic compounds that were used in the correlation. Values are both from DME and liquid ammonia.}
    \begin{tabular}{ccccccc}
         & \multicolumn{3}{c}{DME} & \multicolumn{3}{c}{NH$_3$} \\
         scan rate & $E_\mathrm{pc}$ & $E_\mathrm{i}$ & $E_\mathrm{e}$ & $E_{1/2}$ & $E_\mathrm{pc}$ & $E_\mathrm{e}$ \\
         250~mV/s & pseudo & & & pseudo & & \\
         \midrule
         nitrobenzene & $-1.22$ & $-1.13$ & $-1.00$ & $-0.27$ & $-0.30$ & $-0.22$ \\
         benzophenone & $-1.89$ & $-1.71$ & $-1.53$ & $-1.03$ & $-1.06$ & $-0.98$ \\
         $p$-nitrotoluene & $-1.19$ & $-1.10$ & $-0.93$ & $-0.31$ & $-0.34$ & $-0.27$ \\
         1-nitronaphthalene & $-1.00$ & $-0.94$ & $-0.70$ & $-0.25$ & $-0.29$ & $-0.20$ \\
         4-cyanopyridine & $-1.92$ & $-1.73$ & $-1.54$ & $-1.07$ & $-1.11$ & $-1.02$ \\
    \end{tabular}
    \label{tab:CV-potentials}
\end{table}

\section{Absolute Potential of the SHE}

Note that the Gibbs free energy of solvation of the proton in liquid ammonia, the key part in the calculation of absolute redox potential of the SHE, is significantly temperature dependent.
We have adopted the temperature dependence of  $\Delta G^{(\mathrm{am})}_{\mathrm{H}^+}$ from Ref~\citenum{Malloum2017-10.1063/1.4979568}: 
\begin{equation}\label{eq:E_SHE}
    \Delta G_\mathrm{am} (\mathrm{H}^+, T) = -1265.832 + 0.210 T
\end{equation}
Using the experimental temperature T = 220 K, the final proton solvation energy in liquid ammonia used for the further calculations is  $\Delta G^{(\mathrm{am})}_{\mathrm{H}^+}=-1219.63$~kJ/mol.

\section*{REFERENCES}
%